\documentclass[]{spie}  

\usepackage{amsmath,amsfonts,amssymb}
\usepackage{graphicx}
\usepackage{color}
\usepackage[colorlinks=true, allcolors=blue]{hyperref}

\title{Nonlinear dynamics in multimode optical fibers}

\author[a,b]{Stefan Wabnitz}
\author[c]{Alessandro Tonello}
\author[c]{Vincent Couderc}
\author[a]{Daniele Modotto}
\author[c]{Alain Barth\'el\'emy}
\author[d]{Guy Millot}
\author[a]{Katarzyna Krupa}

\affil[a]{Dipartimento di Ingegneria dell'Informazione, Universit\`a di Brescia, via Branze 38, 25123, Brescia, Italy}
\affil[b]{Istituto Nazionale di Ottica del Consiglio Nazionale delle Ricerche (INO-CNR), Via Branze 45, 25123 Brescia, Italy}
\affil[c]{Universit\'e de Limoges, XLIM, UMR CNRS 7252, 123 Av. A. Thomas, 87060 Limoges, France}
\affil[d]{Universit\'e de Bourgogne, ICB, UMR CNRS 6303, 9 Av. A. Savary, 21078 Dijon, France}

\begin{document} 
\maketitle

\begin{abstract}
 We overview recent advances in the research on spatiotemporal beam shaping in nonlinear multimode optical fibers. An intense light beam coupled to a graded index (GRIN) highly multimode fiber undergoes a series of complex nonlinear processes when its power grows larger. Among them, the lowest threshold effect is the Kerr-induced beam self-cleaning, that redistributes most of the beam energy into a robust bell-shaped beam close to the fundamental mode. At higher powers a series of spectral sidebands is generated, thanks to the phase matching induced by the long period grating due to the periodic self-imaging of the beam and the Kerr effect. Subsequently a broadband and spectrally flat supercontinuum is generated, extending from the visible to the mid-infrared. 
\end{abstract}


\section{INTRODUCTION}
\label{sec:intro}  
Multimode optical fibers have been studied for a long time in the context of fiber communication systems. Their possible 
application to spatial division multiplexing (SDM) was proposed more than thirty years ago \cite{bib:pfacq}, however their large modal dispersion prevented their use in long distance fiber optics links.
Besides telecommunications, the practical use of multimode optical fibers (MMFs) has been quite limited so far, 
because of their inherent inability to maintain a good beam quality, owing to modal scrambling. 
In recent years, there has been a resurgence of interest in the use of MMFs for increasing the transmission capacity of coherent
transmission systems via SDM, thanks to the possibility of digital dispersion compensation. Moreover, the use of of MMFs may permit the scalability of the output power from fiber lasers and fiber based supercontinuum sources. In the context of both optical communications systems and fiber laser sources, nonlinear propagation effects in the MMF provide a fundamental limitation to their stable operation. 

Nevertheless, not until recently it has been fully realized that nonlinear MMFs may exhibit a rich and complex spatiotemporal propagation dynamics\cite{Picozzi2015R30}. 
 For example, for pulses propagating in the anomalous dispersion regime, femtosecond multimode optical solitons can be formed
  in graded-index (GRIN) MMFs \cite{Renninger2012R31}. These temporal wave packets permit to compensate modal dispersion by nonlinearity. Moreover, their spatiotemporal intensity oscillations can be exploited for the controlled generation of ultra-wideband dispersive wave series \cite{WrightNatPhot2015,Wright2015R29}. This work is dedicated to presenting an overview of what perhaps is the most intriguing property of MMFs, namely the possibility of intimately coupling, via the fiber nonlinearity, their spatial (one longitudinal, and two transverse) dimensions with the temporal dimension of guided wave propagation.

\section{KERR BEAM SELF CLEANING UNDER VARIOUS PUMP EXCITATIONS}
\label{sec:sc}

 \begin{figure}[t] 
   \begin{center}
   \begin{tabular}{c} 
    \includegraphics[height=4.5cm]{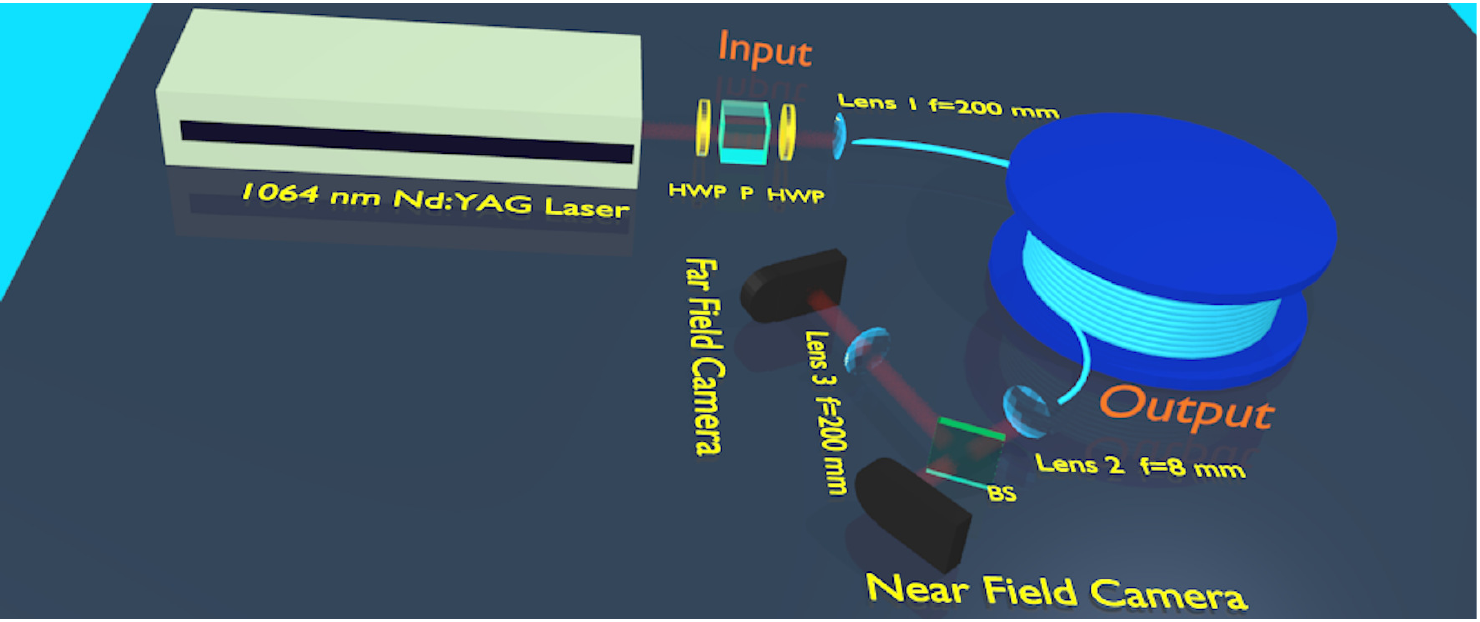}
       \end{tabular}
   \end{center}
   \caption[example] 
   { \label{fig:setup} 
Experimental setup comprising the laser source, the GRIN MMF and the detection system at the output.}
   \end{figure} 

Our first series of experiments of nonlinear effects in GRIN MMFs were carried out with 
a flash-pumped Nd-YAG laser emitting pulses at 1064\,nm with a temporal duration
of 30\,ps and a repetition rate of 20\,Hz. An internal option of the laser permitted also 
to generate pulses at the laser second harmonic of 532\,nm. At those wavelengths the light pulses propagate
in the normal dispersion regime. 
We considered segments of GRIN fiber of different lengths up to 30\,m; 
the fiber was coiled with spirals of about 10\,cm 
in diameter.
We analysed the output beam far field and near field by using two cameras and an example of
 experimental setup is illustrated in fig.\ref{fig:setup} . 

\begin{figure}[ht] 
   \begin{center}
   \begin{tabular}{c} 
      \includegraphics[height=7.5cm]{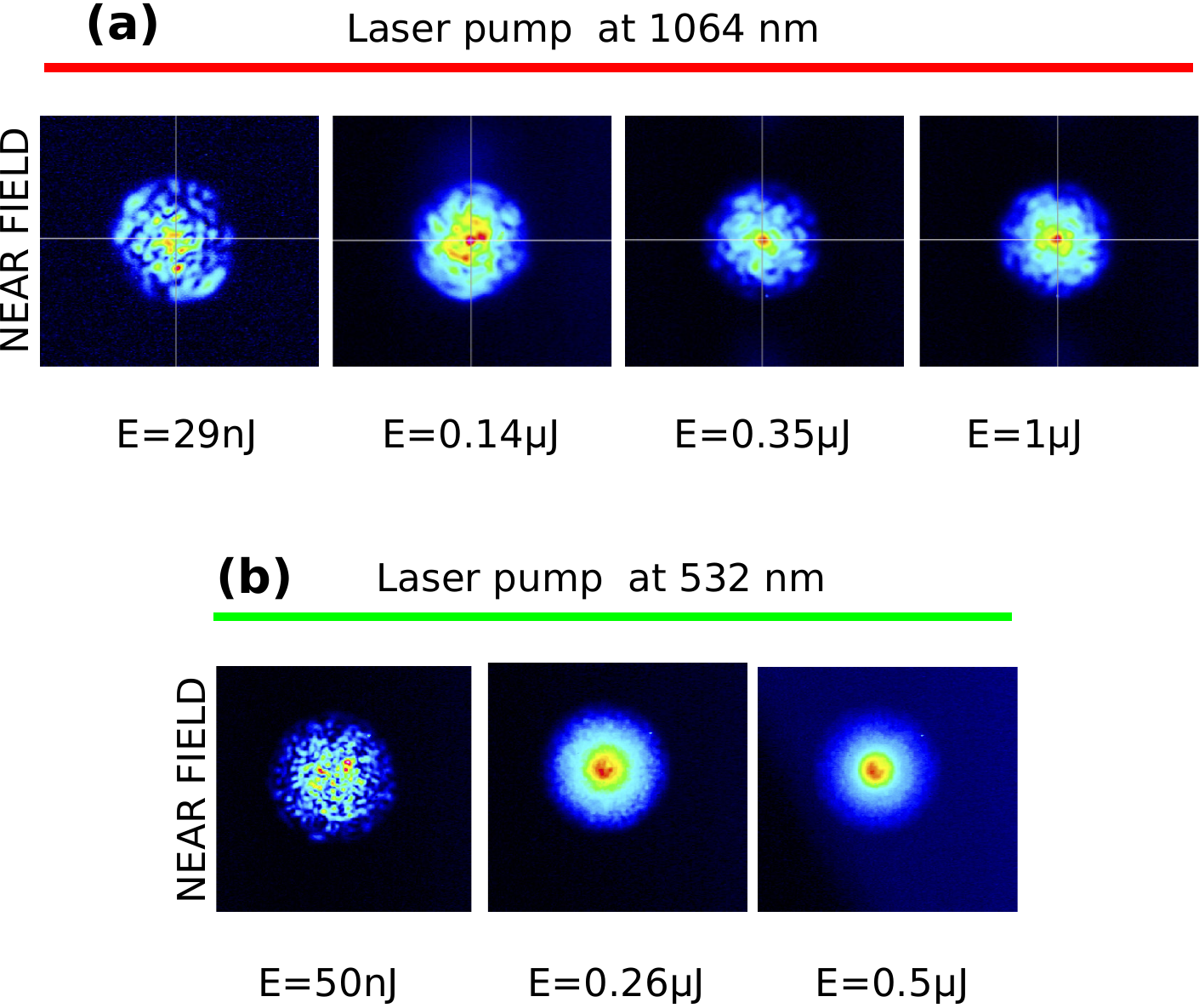}   
      \end{tabular}
   \end{center}
   \caption[] { \label{fig:exp30ps} Experimental observations of beam self cleaning as the input pulse energy grows larger,  
   with pulses of 30\,ps and a laser wavelength of 1064\,nm (a) and  532\,nm (b); fiber length: 30 m.}
   \end{figure}

The principal outcomes of these early observations are summarized in fig.\ref{fig:exp30ps}. Panel (a) 
shows the output beam (near field) reshaping at 1064\,nm, for increasing input pulse energy levels. As can be seen, the results
are different from the typical speckled output pattern that can be observed 
when low energy pulses are launched in a GRIN MMF (this last case is illustrated by  the first frame of panel (a)
where the pulse energy was of only 0.029\,$\mu J$). 
In the linear regime, the speckle sizes are related to the guided modes, 
whose number can be increased by reducing the carrier wavelength. 

Panel (b)  summarizes similar  observations carried out by using the frequency doubled pump at 532\,nm.
The insets at low energy show that the speckle grains are far smaller than that of the previous case at 1064\,nm,
as a natural consequence  of the increased number of guided modes that is obtained by
 halving the pump wavelength.  Nevertheless we observed a similar effect of beam reshaping with an increased brightness
 in the center of the beam. 
The images of the output beam shapes have been obtained by using  a filter at 532\,nm with a bandwidth 
of 10\,nm. Note that in both cases (at 1064 nm and at 532 nm)  the beam self-cleaning has been observed 
also in the far field and  before the appearance of any frequency conversion phenomena, such as Raman Stokes sidebands. 

\begin{figure}[t] 
   \begin{center}
   \begin{tabular}{c} 
      \includegraphics[height=6cm]{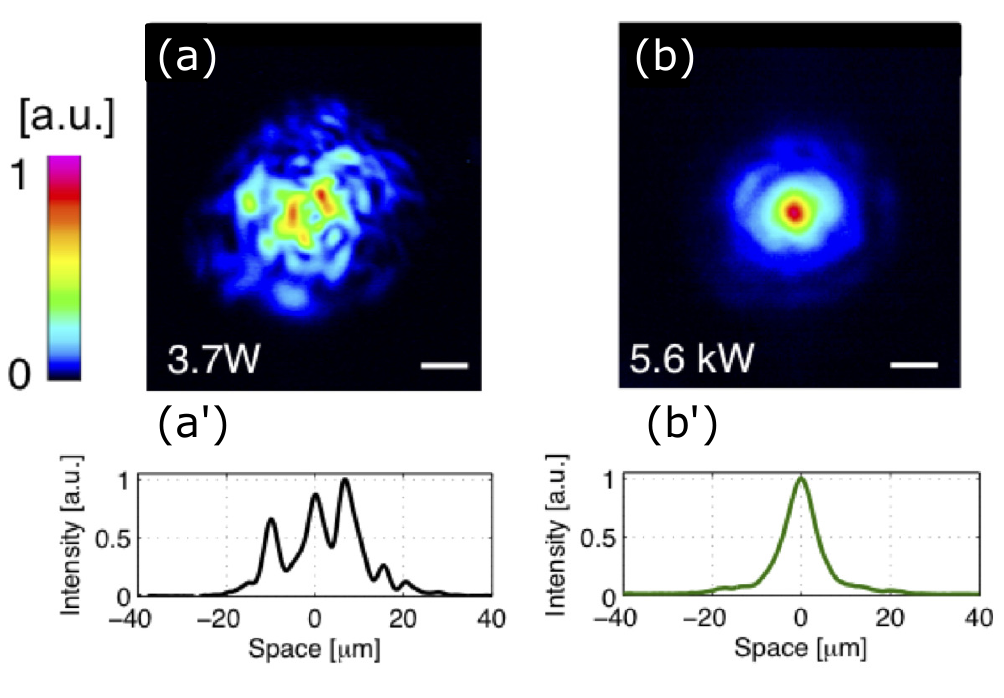}
   \includegraphics[height=6cm]{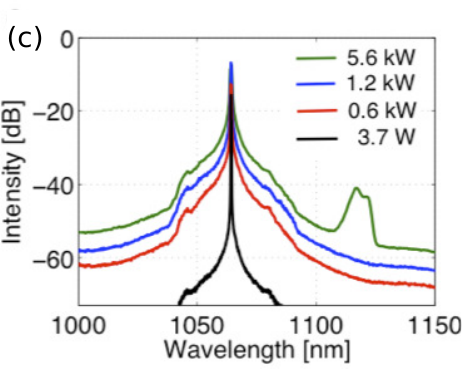}
      \end{tabular}
   \end{center}
   \caption[] {\label{fig:ns} Experimental observations of beams self cleaning with 900\,ps pulses and 30\,kHz repetition rate; fiber length: 12 m.}
   \end{figure}

Next we moved to a different type of laser source based on an amplified microchip  Q-switched 
 Nd-Yag laser, still emitting pulses at 1064\,nm but with a much longer pulse duration of 900\,ps, and with 
a higher repetition rate of 30\,kHz  \cite{Krupanatphotonics,KrupaPRLGPI}. 
 With this second type of laser we confirmed and improved the previous observations 
of a power dependent beam reshaping towards a bell-shaped beam, as can be seen by 
panels (a), (b) of fig. \ref{fig:ns} as well as the corresponding profiles (a') and (b') 
obtained with low (3.7\,W) and high (5.6\,kW) peak power respectively. Panel (c) of the same figure shows how the beam cleaning is obtained nearly in absence of Raman conversion.
Similar results have been also reported by Z. Liu et al. in Ref.\citenum{LiuKerr} using femtosecond pulses.

In another series of experiments we demonstrated effective beam cleaning in an Ytterbium doped fiber. 
The use of an active fiber substantially widens the interest for this effect, which is compatible with the process of
population inversion by using an additional pump \cite{GuenardOpex}.
The principal outcome of our work was that the presence of distributed gain in the active fiber can 
reduce the physical length required for obtaining the beam cleaning. We also extended our 
work to the case of a coupled cavity laser made by 3 mirrors \cite{Guenard:17}: the inner cavity was a microchip 
Q-switched laser, whereas the outer cavity was obtained by a pumped active multimode fiber. 
The large difference in size of the two cavities eased the process of finding common longitudinal modes
in the compound cavity. 

 \section{ GEOMETRIC PARAMETRIC INSTABILITY, SECOND HARMONIC  AND SUPERCONTINUUM GENERERATION IN GRIN MMFs} 
\label{sec:sup} 
Once obtained the beam self-cleaning effect in GRIN MMFs, a further increase in the input power leads to
 the generation of a series of spectral sidebands ranging from the visible to the infrared. This process can be understood as a spatiotemporal instability,
induced by the periodic evolution of the light in a GRIN fiber. This instability occurs both in the normal and in the anomalous dispersion regime of GRIN fibers \cite{Longhi2003R27}. 
Note that in a previous work with short input pulses propagating in the anomalous dispersion regime \cite{Wright2015R29},
 L. G. Wright {\it et al.} discussed the experimental observation of dispersive waves series radiated by multimode solitons: these sidebands result from the same light-induced refractive index periodicity along the longitudinal direction.

Although the GRIN fiber is in itself a longitudinally invariant waveguide, 
light propagation in such a fiber exhibits a periodic evolution (or self-imaging), owing to the parabolic transverse profile of the refractive index. This profile corresponds to a harmonic oscillator potential, leading to equally spaced propagation constants for the linear eigenmodes of the fiber. Consequently, a longitudinal periodic modulation of the refractive index is established, thanks to the periodic overshoots of the light intensity, which results into a local periodic oscillation of the refractive index via the Kerr effect.  Although such long-period grating is extremely shallow, when compared with the transverse refractive index variation of the fiber itself, it leads to important consequences on fiber propagation, owing to the long interaction length.
Note that such  grating is not permanently written in the fiber, as in the case of Bragg gratings, 
but instead is traveling along with the pulse itself, and it vanishes when the input light intensity is reduced to low values. 

The presence of a dynamic or light-induced long-period grating was previously exploited with short segments of GRIN fiber: these were used as mode couplers, whenever the grating wave vector could compensate for the propagation constants of two signal modes propagating at a different wavelength \cite{Hellwig:14}. 
 In our case, the longitudinal index periodicity provides a quasi-phase matching (QPM) for a special type of spatiotemporal instability, that in normal dispersion regime was dubbed as geometric parametric instability (GPI), 
 to emphasize its origin associated with fiber geometry, as opposed to previously known parametric instabilities, which occur in intrinsically periodic structures. 
 
 In a GRIN MMF,
the self-imaging period $\Lambda_S$ is related to the fiber transverse parameters by the relation $\Lambda_S=\pi\rho/\sqrt{2\Delta}$, 
where $\rho$ is the fiber core radius and $\Delta$ is the relative refractive index difference.
For standard GRIN fibers, the self-imaging period is sub-millimetric, and can provide the necessary momentum for quasi-phase matching at several harmonic orders \cite{Wright2015R29,KrupaPRLGPI,Matteo:17}, as it was illustrated by a series of experiments with sub-nanosecond pulses in Ref. \citenum{KrupaPRLGPI}.
In our experiments, we verified that parametric sidebands related to the periodical evolution of light propagation are located at large frequency detuning from the pump, therefore having a weak dependency
upon the quasi-CW beam intensity. This permits to approximate well their frequency offsets 
from the pump as $f_h\simeq\pm\sqrt{h}\Delta f$, where $2\pi \Delta f\simeq\sqrt{2\pi/(\Lambda_S\kappa'')}$. 
In contrast, we may note that the spectral positions of other types of parametric instabilities, which take place in MMFs for relatively small detuning from the pump, do depend on light intensity, as discussed in 
Ref.\citenum{Dupiol:17}.

For the case of GPI, with a standard GRIN MMF with a $\rho=26$ $\mu m$, group velocity dispersion $\kappa^ {''}=16.55\times 10^{-27}s^2/m$ at the pump wavelength of 1064~nm, and refractive index difference $\Delta=8.8\times 10^{-3}$, one can estimate the self-imaging period as $\Lambda_S=0.615$ $mm$. These values lead us to analytically estimate  the frequency detuning for the first resonant sideband, namely $\Delta f\simeq 125 THz$. However its precise value depends on the knowledge of the opto-geometric parameters of the fiber at the pump wavelength. 

 \begin{figure}[t] 
   \begin{center}
   \begin{tabular}{c} 
      \includegraphics[height=5cm]{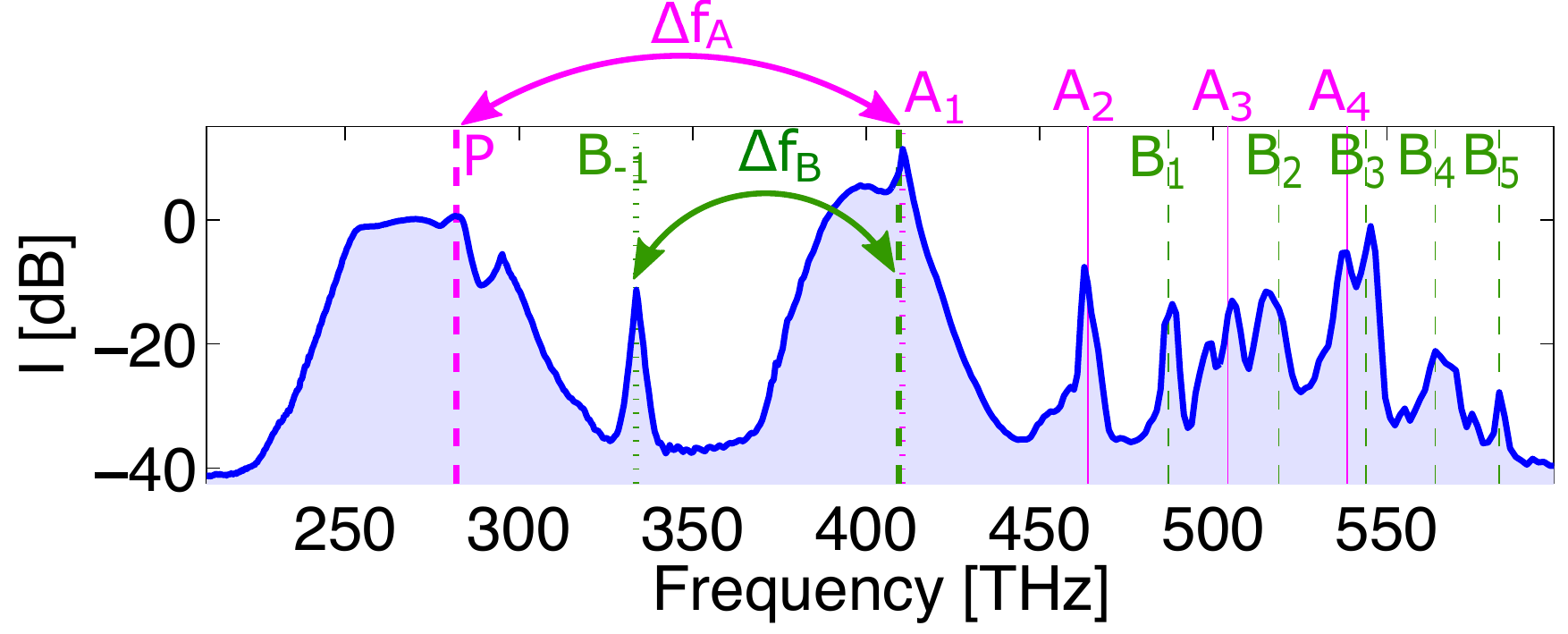}
         \end{tabular}
   \end{center}
   \caption[] { \label{fig:secondarypump} Experimental spectrum obtained from a 3-m long GRIN fiber with a radius $\rho=26\mu m$ and an input peak power of 51.8\,kW.}
   \end{figure} 

We show in fig.\ref{fig:secondarypump} an example of experimental results obtained in a 3-m long 
GRIN fiber, pumped by a sub-nanosecond laser with an input peak power of 51.8\,kW. The spectral position of the pump is indicated by the letter P (note that the spectrum of the residual pump is not well resolved because it is much narrower than the spectral resolution of our OSA). 
To check the self-consistency of our interpretation, 
one can estimate first the frequency shift $\Delta f_A$ between the pump P and the first anti Stokes sideband 
($A_1$) directly from the experiment: $\Delta f_A=128.6 THz$, which agrees fairly well with the value predicted by the simple analytic formulation. Once fixed the frequency detuning of the first sidebands, the other spectral orders are expected to be detuned by multiples $\sqrt{h}\Delta f_A$, where $h$ is an integer number. The
purple solid lines $A_2\dots A_4$ represent these values, and one can check how these spectral positions coincide very well with  local spectral peaks. Another important observation is the
presence of additional peaks besides the already discussed sidebands  $A_2\dots A_4$. This fact can be explained as follows:  from fig. \ref{fig:secondarypump} we can observe a strong conversion efficiency towards the first order anti Stokes sideband $A_1$ at 730.8\,nm (410.5\,THz). Such sideband can in turn generate its own series of sidebands, based on the same process governed by the periodic intensity evolution along the fiber. 

In order to check the consistency of this statement, one can proceed in a similar way: first one can measure, from the experimental results, the spectral detuning $\Delta f_B$ between $A_1$ (the new pump) and the first order Stokes sideband of this new series, indicated as $B_{-1}$, because in this case the first Stokes sideband falls within the spectral window of the experimental results. The experimental value is $\Delta f_B=76.7 THz$.
Note that $\Delta f_B=\Delta f_A/1.67$ and this fact can be explained by the change in group velocity dispersion caused by the change of wavelength. At 730.8\,nm silica glass has
 $\kappa^ {''}=42.13\times 10^{-27}s^2/m$. By increasing the GVD of a factor 2.54 one can explain a reduction in frequency detuning by a factor $\sqrt{2.54}=1.59$ which fits fairly well with the reduction observed of 
 a factor 1.67. 
 
 Note that in this reasoning we assumed an identical self-imaging period at both wavelengths. Again by looking at the series of sidebands spaced by frequency detunings $\sqrt{h}\Delta f_B$ and indicated for ease of comparison by labels $B_1\dots B_5$ we can obtain a perfect agreement with nearly all the remaining peaks. 
 The spectrum of fig. \ref{fig:secondarypump} can be then interpreted as a composition of two series of sidebands:
 the first series is generated by the pump at 1064\,nm; the second series of sidebands is instead generated by the $1^{st}$ order antiStokes, whose conversion was particularly efficient. The possibility to have parametric effects induced by a secondary pump was discussed by R. Dupiol et al. in Ref. \citenum{Dupiol:17fardetuned}.
Such multiple parametric generation has been explained as a form of self-organized instability  by L. G. Wright et al. in Ref.\citenum{WrightNP2016}.

Light induced periodicities are not limited to self-imaging in a GRIN fiber. 
 Another effect that can spontaneously be activated in glass fibers, especially with Germanium doping, is the so called optical poling. 
 In short, the  exposure (few hours) of an optical fiber to an intense laser beam at 1064\,nm may lead to the 
 generation of a photo-induced charge distribution, which in turn gives rise to a permanent modulated 
 equivalent quadratic nonlinearity \cite{Osterberg:87}, 
 and whose period satisfies the QPM condition.  
 The estimated QPM period $\Lambda$ for second harmonic generation (SHG) in glass fiber 
 is of 48\,$\mu m$ for a pump wavelength at 1064\,nm.
 
\begin{figure}[t] 
   \begin{center}
   \begin{tabular}{c} 
      \includegraphics[height=6cm]{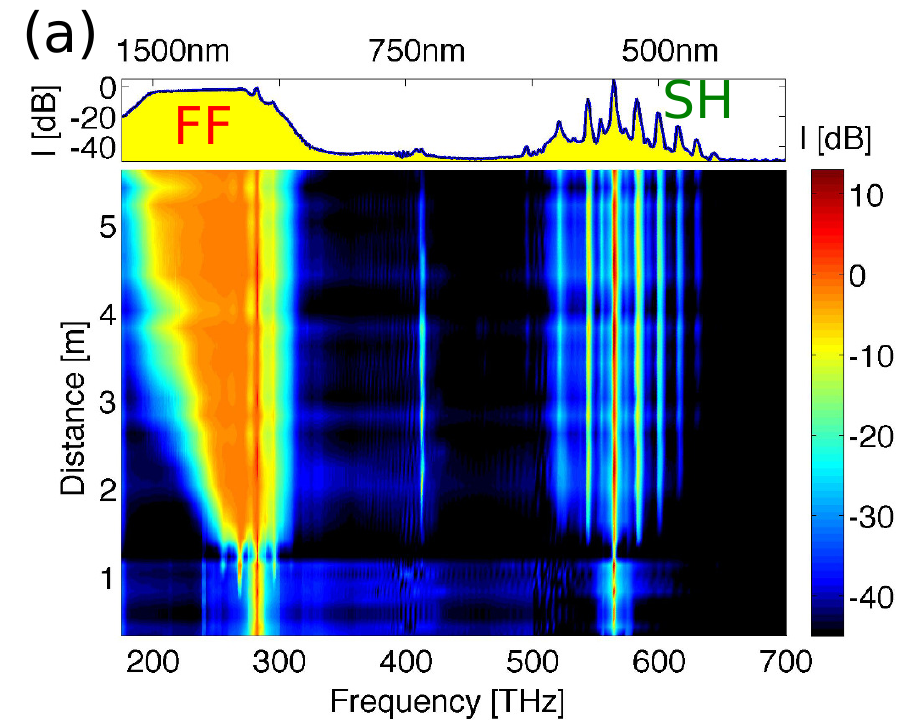}
   \includegraphics[height=5.5cm]{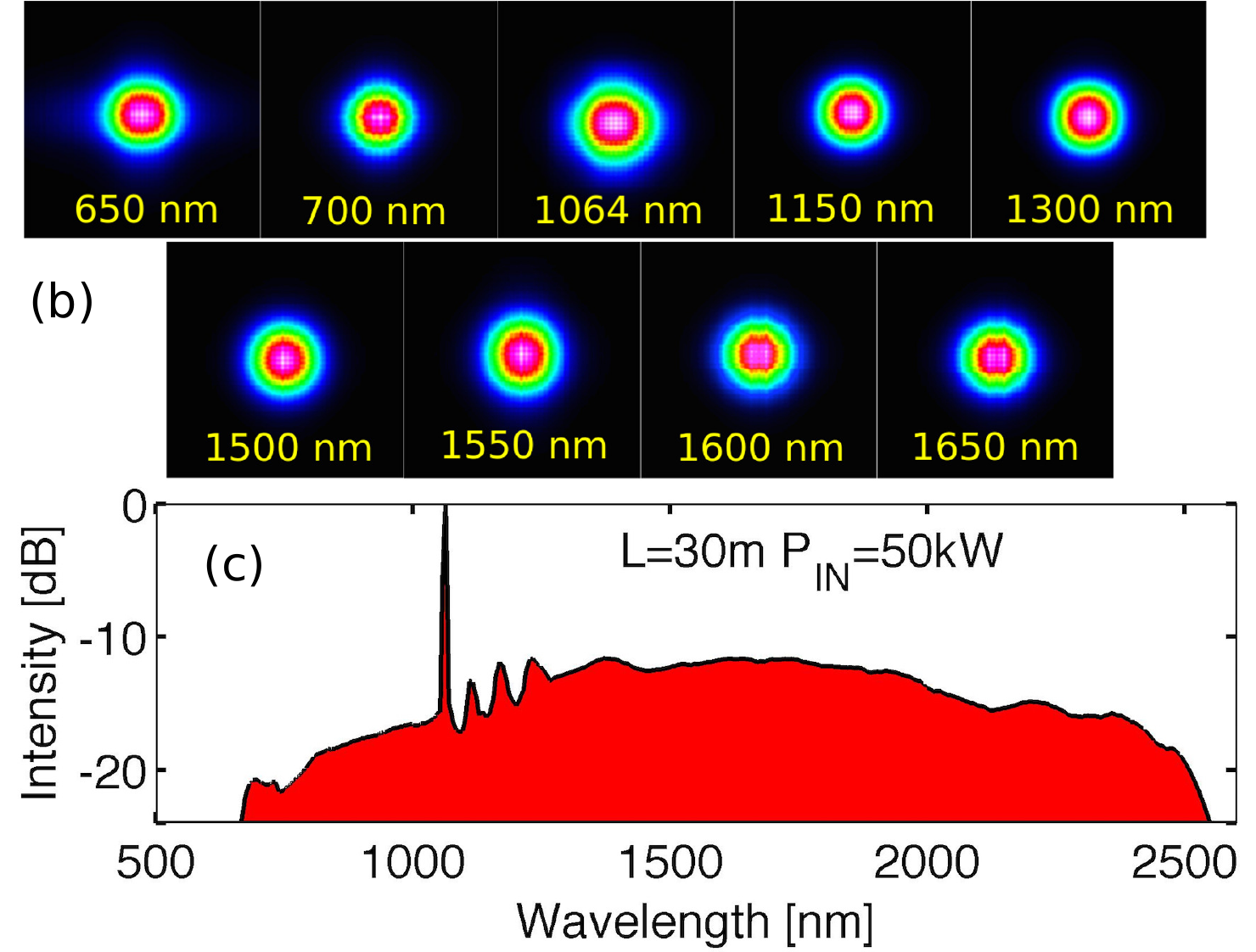}
      \end{tabular}
   \end{center}
   \caption[] { \label{fig:sc} (a) Second harmonic generation in an optically poled GRIN fiber and broadband pump. Spatial (b) and spectral features (c) of a Supercontinuum generation in a 30-m long GRIN fiber with an input peak power of  50\,kW}
   \end{figure}

 Interesting features can happen in optically poled GRIN fibers, owing to the presence of these
 two periodicities of different scales \cite{CeoldoSHG}. 
 A synthesis of the experimental results is given in panel (a) of fig.\ref{fig:sc}. In particular it is possible to observe how in the first meter of propagation, a narrowband pump generates a narrowband SH field. As long as 
 the pump spectrum broadens, for instance due to the Raman effect, the spectrum at the SH becomes populated by sharp unevenly spaced peaks.  
 
 To explain the experimental results, we can consider that in an optically poled MM-GRIN fiber, 
 the efficiency of the poling process is influenced by the self-imaging effect. 
 In fact, the photo-induced charge redistribution is enhanced where the light intensity is higher.
The consequent additional periodic variation of the nonlinear coefficient leads to the following 
modified QPM for poled MM-GRIN fibers

\begin{equation}
\beta(2\omega)-2\beta(\omega)-\frac{2\pi}{\Lambda}-q\frac{2\pi}{\Lambda_S}=0
\label{eq:1}
\end{equation}
where  $\beta(\omega)$ ($\beta(2\omega)$) is the linear propagation constant of the pump (SH field), and $\Lambda$ is the QPM period for a monochromatic pump at 1064\,nm. 
 $q=0,\pm 1, \pm 2, \dots$ is associated with each different harmonic of the self-imaging period, and is physically related to the sharp peaks observed in the experiments around the main SH peak at 532 nm.
The amount of momentum contribution given by $\Lambda_S$ is in fact 
nearly one order of magnitude smaller than the momentum associated with $\Lambda$.

Graded index fibers are also interesting waveguides for supercontinuum generation in the visible and near infrared as it 
was also pointed out by G. Lopez-Galmiche et al. in Ref. \citenum{Galmiche:16}.
The supercontinuum generation inherits the bell shaped spatial profiles discussed for the 
multiple parametric sidebands \cite{KrupaPRLGPI}. We also analyzed in this context the spatio-temporal
features of the supercontinuum generation unveiling the different arrival time and pulse duration
under various wavelength windows  \cite{KrupaLuot:16}. We showed how a judicious 
choice of the fiber length reduces the spectral ripples due to the Raman conversion.
A selection of our results is presented in panels (b) and (c) of fig.\ref{fig:sc}.

 \section{CONCLUSIONS} 
A series of recent experiments have shown that the nonlinear response of highly multimode fibers may be exploited for achieving the intensity controlled shaping of the spatial beam at their output. This remarkable finding permits to effectively counter the natural tendency of multimode fibers to scramble the spatial degrees of freedom of light, which has the potential for enabling a wide range of novel applications. We have demonstrated the generation of spatially quasi single mode octave spanning supercontinuum from a GRIN MMF, and effective second harmonic generation after optical poling. Spatial beam self-cleaning is enhanced in active fiber amplifiers, thus it is expected to provide a key building block for a new class of high power fiber lasers \cite{SpationtempMLWright,GuenardOpex}.

\acknowledgments 
 
We acknowledge the financial support of: Horiba Medical and BPI france within the Dat@diag project; 
 iXcore research foundation; French National Research Agency ANR Labex ACTION; the European Research Council (ERC) under the European Union's Horizon 2020 research and innovation programme (grant agreement No.~740355);
the European Union's Horizon 2020 research and innovation programme under the Marie~Sk\l{}odowska-Curie grant agreement No.~2015-713694 ("BECLEAN" project). 

\bibliography{CleanSpiePWest.bib} 
\bibliographystyle{spiebib} 

\end{document}